\documentclass[aps,floatfix,showpacs]{revtex4}
\def\beq{\begin{equation}}
\def\eeq{\end{equation}}
\def\bea{\begin{eqnarray}}
\def\eea{\end{eqnarray}}
\usepackage{graphicx}

\begin{document}
\title{Superfluid Pairing in the Three Component Fermi Gas}
\stepcounter{mpfootnote}
\author{S. Y. Chang} 
\address{Institute for Theoretical Physics,
		University of Innsbruck, Technikerstr. 25, A-6020 Innsbruck, Austria}
\address{Institute for Quantum Optics and Quantum Information,
		ICT-Geb\"aude, Technikerstr. 21a , A-6020 Innsbruck, Austria}
\author{V. R. Pandharipande}
\address{Department of Physics, University of Illinois at Urbana-Champaign,
        1110 W. Green St., Urbana, IL 61801, USA}
\date{\today}

\begin{abstract}
 We present an analysis of the SU(3) symmetric model of the strongly interacting three component Fermi gas in the continuum space using quantum Monte Carlo techniques. 
Three body effects predominate in the regime of interaction strength beyond that of threshold of the three particle bound state.
However,  we find that there is an interval of the interaction strength where the 
 SU(2)$\otimes$U(1) broken symmetry superfluidity is possible. For a strong enough
interaction, the SU(3) symmetry is restored and the superfluidity is suppressed.  Within the interval of the
broken symmetry, we also find that on average the particle pairs belonging to the species with superfluid pairing remain further
 separated than those without the superfluid pairing correlation. 

\end{abstract}
%\submitto{\NJP}
\maketitle
\section{Introduction}\label{sec_intro}

 A pair of fermions occupying different internal states can interact in s-wave scattering.
Fermi atoms can be loaded into different internal spin projection states to produce
interacting degenerate Fermi gas with $T \approx 0$. Comprehensive reviews of the recent theoretical and experimental advances in the
study of the dilute Fermi gases can be found in the Ref. \cite{giorgini2007,ketterle2008}.
The fermionic alkali atoms can occupy hyperfine states that result from the coupling of the nuclear
angular moment $I$ with the electronic spin $S$. In the presence of a strong magnetic field,
the electronic degree of freedom is polarized and the relevant internal degrees of freedom are determined by the state
 of the nuclear angular momentum. In the case of $^6$Li, the nuclear angular momentum $I=1$ allows three different
projections. By combining magnetic as well as optical trapping fields,
 constraints on the trappable hyperfine states are largely removed.

For a Fermi gas at low density, an expansion of the ground state energy in terms of a small parameter $ak_F$ 
is known \cite{lenz1929,huang1957,galitskii1958}. In general, when we have a gas of fermions occupying $s$ different states with the same partial densities,
we call it the degeneracy $s$ Fermi gas. The ground state energy has an expansion in powers of $ak_F$
%\begin{widetext}
\beq
\frac{E_0}{N} = \epsilon_F\left\{ \frac{3}{5} + 
(s-1) \left[\frac{2 ak_F}{3\pi} + \frac{4}{35\pi^2}(11-2ln2)(ak_F)^2 \right] + {\cal O}[(ak_F)^3] \right\}
\label{eqn_lenz}
\eeq
%\end{widetext}
where we identify the zeroth order term as the free Fermi gas energy $E_{FG} = \frac{3}{5} \epsilon_F$.
 $\epsilon_F = \frac{\hbar^2 k_F^2}{2m}$ is the Fermi energy, $k_F$ the Fermi momentum and $a$ 
the s-wave scattering length. $E_{FG}$ is independent of $s$ and it is used as the unit of energy
throughout this article. The terms that depend on the interaction potential range $R$ are
 to appear at higher orders \cite{baker1999, bulgac2002}. 
For $s = 2$ Fermi gas, the dependence on $R$ can be eliminated by taking the limit $R/r_0 \rightarrow 0$, where $r_0$
 is the average inter-particle distance ($\frac{4}{3}\pi r_0^3 \rho =1$ with $\rho =$ density). 
The {\it intermediate regime} where $R << r_0 \sim \frac{1}{k_F}  << |a|$ is of 
particular interest for $s=2$ Fermi gas.  This is also called the {\it unitarity regime}. 
In this regime, the mean free path of the atoms $\lambda \equiv \frac{1}{\rho \sigma}$ becomes much 
shorter than $r_0$ as the cross section $\sigma$ diverges.  Examples of the systems in this regime are found in the
dilute gases of $^{6}$Li and $^{40}$K atoms at Feshbach resonances close to the zero temperature. 
The magnitude of the s-wave scattering length $|a|$ can be $\sim 1000 \mathring{A}$ at the Feshbach 
resonance while the interaction range of the van der Waals forces $R \sim 10-100 \mathring{A}$. Another example is that of the neutron gas. 
The neutron-neutron interactions by strong force have $a \sim -18.8 fm$ while $ R \sim 1 fm$. 
At $a \rightarrow \pm\infty$, zero energy two particle bound state appears. However, it is known that the 
ground state of the many body system has a positive energy per particle 
$E_0/(N E_{FG}) \approx 0.40 \sim 0.44 $ \cite{carlson2003,carlson2008}.
 Ground state properties for $s = 2$ Fermi gas were studied\cite{carlson2003,astra2004,chang2004} 
using the Quantum Monte Carlo for the unitarity as well as other regimes of interaction. 
The generalization of the low density expansion (Eq. \ref{eqn_lenz}) to $s \ge 3$ becomes troublesome as
the limit of $R \rightarrow 0$ cannot be taken. This is expected since for the three particle systems
 the Efimov effect predicts an effective three particle attractive interaction \cite{bulgac2002} of the form
 $\sim -\frac{s_0^2 \hbar^2}{2mR^2}$ where $s_0$ is a universal constant. Thus, the minimum set of parameters
 to describe the $s=3$ system consist of $ak_F$ and $Rk_F$.
 In general, the mean field treatments of  the three component Fermi gas \cite{honerkamp2004,torma2006,torma2007}
 do not account correctly the three particle physics. 

 Recently, a stable
three component degenerate Fermi gas \cite{jochim2008} has been created experimentally.
 Broad and close lying Feshbach resonances make the strong and
attractive interactions among the atoms in the different internal states possible \cite{gupta2003}.
For simplicity, we label the atoms occupying these three different states by a coloring scheme: {\it Green} (G), {\it Red} (R)  and {\it Blue} (B).
The channel dependent s-wave scattering lengths ($a_{GR}$, $a_{GB}$, and $a_{RB}$) have been measured experimentally \cite{barter2005}. 
It is experimentally difficult to achieve simultaneously strong interactions in all of the channels. 
However,  in the present work we assume a simplified SU(3) symmetric model where $a = a_{GR} = a_{GB} = a_{RB}$.
Also, the mass is assumed to be the same for all the components. We leave non-SU(3) symmetric cases to future study. Three component Fermi gas problem
is also relevant in relation to the {\it color superconductivity} of the quark matter \cite{alford2001}. 

In this article, {\it ab initio} Monte Carlo 
results of the three component Fermi gas are presented. The fermions interact pairwise, by a short but finite range
attractive interaction potential.  The possibility of the SU(2)$\otimes$U(1) broken symmetry ground state \cite{modawi1997, honerkamp2004} 
is considered. Here, only two components participate in the superfluid pairing while the third component remains in the normal phase. 
At weak interaction strengths, the superfluid pairing is exponentially suppressed with the quasiparticle gap $\Delta/T_F \sim e^{\pi/(2ak_F)}$. 
 However, we find that there is an interval of the interaction strength where the broken symmetry pairing has noticeable effects 
in the ground state energy, the quasiparticle gap, the pair distribution functions, etc. When the strength of the interaction is further increased, 
the SU(3) symmetry is restored by the predominant three body effects.   We present our analysis using the dimensional
arguments for the three particle system (Sec. \ref{sec_bound}) and the quantum Monte Carlo method for the three component Fermi gas
(Sec. \ref{sec_many},\ref{sec_results}).

\section{Three Particle Bound State and Scaling Behavior}\label{sec_bound}

L.H. Thomas \cite{thomas1935} noted that in the nucleus of tritium ($^3$H) which has two neutrons and one proton,
the binding energy has no lower bound if we assumed finite negative s-wave scattering length for the proton-neutron
interaction and took the interaction range to zero. Thus, three particle binding energies depend on the range of the potential 
and the ground state energy diverges in the limit of zero interaction range. In addition, as the pairwise interaction approaches 
the resonance ($a \rightarrow -\infty$), infinite number of shallow three particle bound states (called trimers or trions) appear one after 
another. They are known as the Efimov states \cite{efimov70,efimov71,lim77,esry2005}. This property is 
 strikingly different from the two particle case. Consequently, qualitatively different behavior of the three component
 gas is expected to emerge in comparison with the two component Fermi gas.  For our discussion, we consider a generic
 three particle Hamiltonian where the particles interact pairwise
\beq
{\cal H}_3 = -  \frac{\hbar^2}{2m} \sum\limits_{i \in \{G,R,B\} } \nabla_i^2 + v_0\sum\limits_{i<j \in\{G,R,B\}} V_R(r_{ij})~.
\label{eqn_3body_h}
\eeq
There are two positive parameters in the Hamiltonian: strength $v_0$ and range $R$ of the interaction. 
We assume a negative dimensionless function $V_R(r) \le 0$ while $v_0 >0$ has dimension $\sim length^{-2}$. $V_R(r)$ is solely parametrized by $R$. 
The s-wave scattering length $a$ can be used equivalently instead of the strength $v_0$.

In the two particle systems the bound state threshold is at the resonance ($a_2^c = \pm \infty$) independent of the finite $R$.
 Given a value of $R$, it is easy to see that three particle threshold $a_3^c$ can be such that $a_3^c< 0$ but not equal to $-\infty$. 
This can be done using an ansatz as the trial wave function; for example, a function of the form $\Psi_{3,trial}({\bf R}) = f(r_{GR}) f(r_{GB}) f(r_{RB})$
with the variational $f(r)$.  In this case, it suffices to provide a trimer state upper bound in energy to prove the existence of a trimer state with
$a_3^c$ in the interval $(-\infty,0)$ for a given $R$.  The fact that $a_3^c$ is negative and $a_3^c \ne -\infty$ is the starting point for 
the analysis on the scaling behavior of the length unit. When we consider
 rescaling of the length by taking  $ R \rightarrow \alpha R$ and $a \rightarrow \alpha a$ (we will always assume $0 < \alpha <1$ from now and on),
 consistent scaling behavior with $ \langle {\cal H}_3 \rangle \rightarrow \langle {\cal H}_3 \rangle / \alpha^2 $ 
is expected (we will justify in the next paragraph). Here, it can be seen easily from the zero energy scattering solution that the s-wave scattering length $a$ scales analogous 
to another length quantity $R$. By making  $\alpha \rightarrow 0^+$ we expect $R \rightarrow 0^+$, $a_3^c \rightarrow 0^-$ and  $\langle {\cal H}_3 \rangle  \rightarrow -\infty$ 
for any $a$ that belongs to the interval $(-\infty, a_3^c \rightarrow 0^-)$. This means that, for a zero range interaction, 
the trimer state is possible for any attractive pairwise interaction of nonzero strength.

Now, we consider formally the scaling behavior of the energy for the system of three particles in vacuum.
We assume a value of $R$ such that trimer state is allowed with a strength $v_0^c$ corresponding to $a_3^c < 0$. 
The three particle Schr\"odinger equation with the usual notation ${\bf X} = \{ {\bf x}_G,
{\bf x}_R, {\bf x}_B \}$ and $r_{ij} = |{\bf x}_i - {\bf x}_j|$ is
\beq
 {\cal H}_3 \Psi({\bf X}) = E \Psi({\bf X}) ~.
\label{eqn_3body_sch}
\eeq
Let ${\bf X}_s = \alpha {\bf X}$, $R_s = \alpha R$, $v_{0,s} = v_0/\alpha^2$, $r_{ij,s} = \alpha r_{ij}$  and $\Psi_s({\bf X}_s) = \Psi({\bf X})$. 
Then $\nabla_i^2 \Psi_s({\bf X}_s) =\frac{1}{\alpha^2} \nabla_i^2 \Psi({\bf X})$ and 
$V_{R_s}(r_{ij,s}) \Psi_s({\bf X}_s) = V_R( r_{ij}) \Psi({\bf X})$.
Thus, after scaling all the length quantities by $\alpha$, the Eq. \ref{eqn_3body_sch} becomes
\bea
& &\left[- \frac{\hbar^2}{2m} \sum_{i \in \{G,R,B\}} \nabla_i^2 + 
 v_{0,s} \sum_{i<j \in \{G,R,B\}} V_{R_s}(r_{ij,s})
          \right] \Psi_s({\bf X}_s)  \nonumber \\
& = &\left[- \frac{1}{\alpha^2}\frac{\hbar^2}{2m}\sum_{i \in \{G,R,B\}}  \nabla_i^2 + \frac{v_0}{\alpha^2} \sum_{i<j\in \{G,R,B\}} V_R(r_{ij})  \right] \Psi({\bf X}) \nonumber  \\
& = & \frac{1}{\alpha^2}E \Psi({\bf X})= E_s \Psi_s({\bf X}_s) ~.
\eea
As a result, after length scaling  $\Psi_s({\bf X}_s)$ is the solution with the eigenvalue $E_s = \frac{E}{\alpha^2}$. 
We arrived at this property by using the dimensional arguments alone.
As good illustrative examples of this scaling behavior, the following cases are mentioned:
  When $a \in (-\infty, a_3^c)$ for a given $R$, there is trimer state with energy $E<0$. 
Then, we can scale length by an overall factor $\alpha$ but keep $a_s$ constant (that is, increase $|a_s|$ to match $|a|$). 
It is obvious that taking $\alpha \rightarrow 0^+$ causes $E_s(a)$ to collapse rapidly to $-\infty$. 
This means that given a trimer state in vacuum, when $R \rightarrow 0$ at a fixed non-zero value of $a$ ($<0$), $E$ has to go to $-\infty$.
This is in agreement with the above mentioned Ref. \cite{thomas1935}. As another example, if we had initially
 $a \in (a^c_3,0)$, the interaction potential is not strong enough to allow trimer state and the total energy of the particles $E = 0$ in the vacuum. 
 In this case, length scaling leaves the particles unbound with $E_s(a_s) = 0$. Thus, unbound particles remain unbound even after length scaling
and the collapse does not occur. 

For the comparison purpose, let's also consider the scaling behavior of a pair of particles. We can see that
the scaling of length does not produce the collapse as in the three particle case. For the s-wave scattering length $a$
such that $1/a < 0$, there is no bound state and the energy of the pair remains zero ($E = 0$) in the vacuum at any length scale. 
Then, we consider $1/a > 0$ regime (usually called BEC regime). Using the same scaling analysis, we also have $E_s(a_s) = \frac{E(a)}{\alpha^2}$.
 In this case, we can solve exactly the contact interaction ($R=0$) problem by replacing the potential by the boundary condition $\frac{u'(0)}{u(0)} = -\frac{1}{a}$ where $u(r)$ is the radial 
wave function of the pair. The solution for the radial wave function is $u(r) \sim e^{-\frac{r}{a}}$ with 
$E_{pair}(a) = -\frac{\hbar^2}{m a^2}$. This energy is finite unless $a \rightarrow 0^+$. According to this solution,
 the scaling behavior of energy $\frac{E_{pair}(\alpha a) }{E_{pair}(a)} = \frac{1}{\alpha^2}$
is the same as the result we obtained from the dimensional analysis, $\frac{E_s(\alpha a)}{E(a)} = \frac{1}{\alpha^2}$.
Thus, for a pair the energy scaling relation becomes identical to the bound state energy in the strong coupling limit $1/a_s \rightarrow +\infty$. 
This is qualitatively different from the three body collapse at the weak coupling limit $1/a_s \rightarrow -\infty$ because
of the simultaneous scaling of $R$ to zero.

\section{Many Particle Ground State}\label{sec_many}

For the study of many body systems at finite density, we use {\it ab~initio} stochastic method known as Fixed Node
Green's Function Monte Carlo (FN GFMC). In general, we take a trial wave function $\Psi_V$ that obeys antisymmetry upon 
the exchange of identical fermions and let it evolve in the imaginary time restricted to a definite
 sign domain given by the nodal surface of the trial wave function itself. If the nodal structure is correct, 
we get the exact ground state. Otherwise, we get an approximate ground state and an energy upper-bound. 
The implementation of this method is explained in detail elsewhere  \cite{carlson2003, chang2004}.

We consider a system with $6 \sim 8$ particles of each color (Green,Red,Blue), so that $18 \le N_{total} \le 24$. 
The particles are contained in a finite box with the periodic boundary conditions at the walls to simulate 
the uniform matter. For many particle systems, we cannot rescale the length without changing the density.  
The scaling behavior analyzed in the previous section is only applicable to the few particle systems in vacuum. 
 In the degeneracy two case, one dimensionless product $a k_F$ uniquely determines the system. The parameter $R$ 
can be pushed in principle to the zero limit and eliminated from the description of the system. In practice,
 small but finite values of $R/r_0 << 1$ were assumed \cite{carlson2003, chang2004} as long as
the results were converged within the statistical errors to the $R k_F \rightarrow 0$ limit. However, from the scaling 
behavior analysis of the previous section, it becomes clear that for the degeneracy three Fermi gas we need to keep
$Rk_F$ finite in order to avoid local trimer instability. Thus, we need both $ak_F$ and $Rk_F$ to fully describe the three component
Fermi gas. We keep $R k_F = 0.32 $
  in the $s = 3$ case which is the same value as in the $s =2$ case ($R/r_0 \sim 0.1$, thus converged to the $R/r_0 \rightarrow 0$ limit).
 However, we should keep in mind that for $s=3$, this particular value of $R$ is not the limit of $R \rightarrow 0$. 
We consider the cases with small deviations from the balanced partial densities.
 We explore the possibility of the superfluid ground state within a certain interval of the interaction strength. 
We also analyze the qualitative behavior of the pair distribution functions (see Sec.\ref{sec_results}). The many body SU(3) symmetric Hamiltonian is 
\beq
{\cal H} = -\frac{\hbar^2}{2m} \sum\limits_{i} \nabla_i^2 +
v_0 \sum\limits_{i<j} V_R(r_{ij})(1-\delta_{c_i,c_j})
\eeq
where $c_i$ is the color of i-th particle. Only pairs of different color particles interact.
$V_R(r)$ is the dimensionless core of the P\"oschl-Teller potential: $V_R(r) = -\frac{1}{ \cosh^2(2 r/R)}$
and $v_0 = \frac{8\hbar^2}{mR^2}$. $v_0$ can be adjusted to get the desired scattering lengths. 
We can impose different nodal restrictions to the solution by using different trial wave functions. We can estimate the energy by using
 SU(3) symmetric Slater trial wave function
\beq
\Psi_{FG}= \Psi_{FG,G}\Psi_{FG,R}\Psi_{FG,B}
\label{eqn_fg}
\eeq
 where the factors represent the normal
 states (given by Slater determinants of the plane wave orbitals) of different
color species. It was also shown\cite{honerkamp2004} that the pairing fields $(\Delta_{GR}, \Delta_{GB}, \Delta_{RB})$
with $\Delta_{\alpha\beta} \sim \sum_{\bf k} \langle c_{{\bf k},\alpha} c_{-{\bf k},\beta}\rangle$ can be mapped
into $(\Delta_0, 0,0)$ with the constraint $\sum_{\alpha,\beta} \Delta_{\alpha\beta}^2 = \Delta_0^2$.
This is analogous to the analysis of Ref. \cite{modawi1997} by Modawi and Leggett where the ground state should allow
one normal phase component. Thus, we consider the  SU(2)$\otimes$ U(1) broken symmetry  pairing ground state with BCS pairing for two of the Fermi components while
the third component remains in the normal phase. The corresponding nodal structure
is given by the trial wave function
\bea
\Psi_{bs-BCS} & = & \Psi_{FG,B} \Psi_{BCS,GR} \nonumber \\
   &=& \left[ \prod_{|{\bf k}|<k_F}a^\dagger_{{\bf k},B} \right] \left[ \prod_{\bf k} (u_{\bf k} + v_{\bf k} a^{\dagger}_{{\bf
   k},G} a^{\dagger}_{-{\bf k},R}) \right]|0\rangle \nonumber \\
    & \rightarrow & \Psi_{FG,B} {{\cal A}}[\phi(r_{11'}) \phi(r_{22'}) ... \phi(r_{MM'})]_{GR}~ .
\label{eqn_bs_bcs}
\eea
Here, we assume that blue species remains in the normal phase (represented by $\Psi_{FG,B}$), while between the green and
red species there is pairing correlation (represented by $\Psi_{BCS,GR}$). In the last line of Eq. \ref{eqn_bs_bcs},
we assumed fixed number projection of the green and red particles.

The complete trial wave function with Jastrow-like factor can be written as
\beq
\Psi_{V, FG} = \prod_{i<j}\left[ f_S(r_{ij})\delta_{c_i,c_j} + f_D(r_{ij})(1-\delta_{c_i,c_j})\right] \Psi_{FG}
\label{eqn_res3_slater}
\eeq
for the SU(3) symmetric Slater wave function (Eq. \ref{eqn_fg}). Here, we considered separately the correlation that exists 
between the same color particles $f_S(r)$ ($ = f_{GG}(r) = f_{RR}(r) = f_{BB}(r)$) from the correlation that exists 
between different color particles $f_D(r)$ ($ = f_{GR}(r) = f_{GB}(r) = f_{RB}(r) $). 
Usually $f_S(r)$ is qualitatively different from $f_D(r)$. $f_S(r)$ is analogous
 to $f_{\uparrow \uparrow}(r)$ of the degeneracy two case and includes Pauli exclusion principle ($f_S(0) = 0$).
 The particular shapes of the correlation functions do not change the GFMC energies. However, 
they are optimized in order to minimize statistical errors and to have the converged pair 
distribution functions $g(r)\approx g_{trial}(r) \approx g_{GFMC}(r)$ (see Ref. \cite{chang2005}).
We can see that the optimized $f_D(r_{ij})$ deviates largely (more peaked at $r \approx 0$) from the one obtained
 by using the LOCV equations \cite{carlson2003}. This is due to the strong three body 
effects even when the pairwise interactions are relatively weak.

Analogously, the complete trial wave function with the SU(2)$\otimes$U(1) broken symmetry pairing correlation (Eq. \ref{eqn_bs_bcs}) is
\bea
\Psi_{V,bs-BCS} & = & \prod_{i<j} [f_S(r_{ij})\delta_{c_i,c_j} + f_{GR}(r_{ij})\delta_{c_i,G}\delta_{c_j,R} +\ldots \nonumber \\
& &  f_{GB}(r_{ij})(\delta_{c_i,G} + \delta_{c_i,R})\delta_{c_j,B} ] \Psi_{bs-BCS}~. 
\label{eqn_res3_bcs}
\eea
This wave function can give better pair distribution functions as the optimization of the correlation functions can be carried out 
separately for $f_{GR}(r)$ and $f_{GB}(r) = f_{RB}(r)$. Then, we can demonstrate that $f_{GR}(0) < f_{GB}(0)$ as well 
as $g_{GR}(0) < g_{GB}(0)$ (see the discussion of the Fig. \ref{fig_two} in the Sec. \ref{sec_results}).
The optimum nodal structure is tried as $\alpha_I= \{0.2, 0.1, 0.02, 0.01, 0.01\}$ with non zero short range
function $\tilde{\beta}(r)$ parametrized by $b=0.44$ (see Ref. \cite{carlson2003,chang2004} for the definitions of these
parameters). These parameters are identical 
to those of the $s=2$ case when $1/ak_F \ge 0$ (called molecular or BEC regime). This trial wave function assumes arbitrarily
 one of the three possible broken symmetry pairing states 
\bea
|1\rangle & \equiv & \Psi_{FG,B} \Psi_{BCS,GR} \nonumber \\
|2\rangle & \equiv & \Psi_{FG,R} \Psi_{BCS,GB} \nonumber \\
|3\rangle & \equiv & \Psi_{FG,G} \Psi_{BCS,RB}.
\label{eqn_degens}
\eea
 These states are degenerate in energy and the broken symmetry can be assumed by choosing one of these states without
loss of generality. The GFMC energies using the trial nodes given by the wave functions of Eq. \ref{eqn_res3_slater} and Eq. \ref{eqn_res3_bcs} 
are summarized in the Fig. \ref{fig_one}, while the pair correlation functions are presented in the Fig. \ref{fig_two}.

\section{Discussion of the Results}\label{sec_results}

\begin{figure}
\begin{center}
\includegraphics[width=8.0cm]{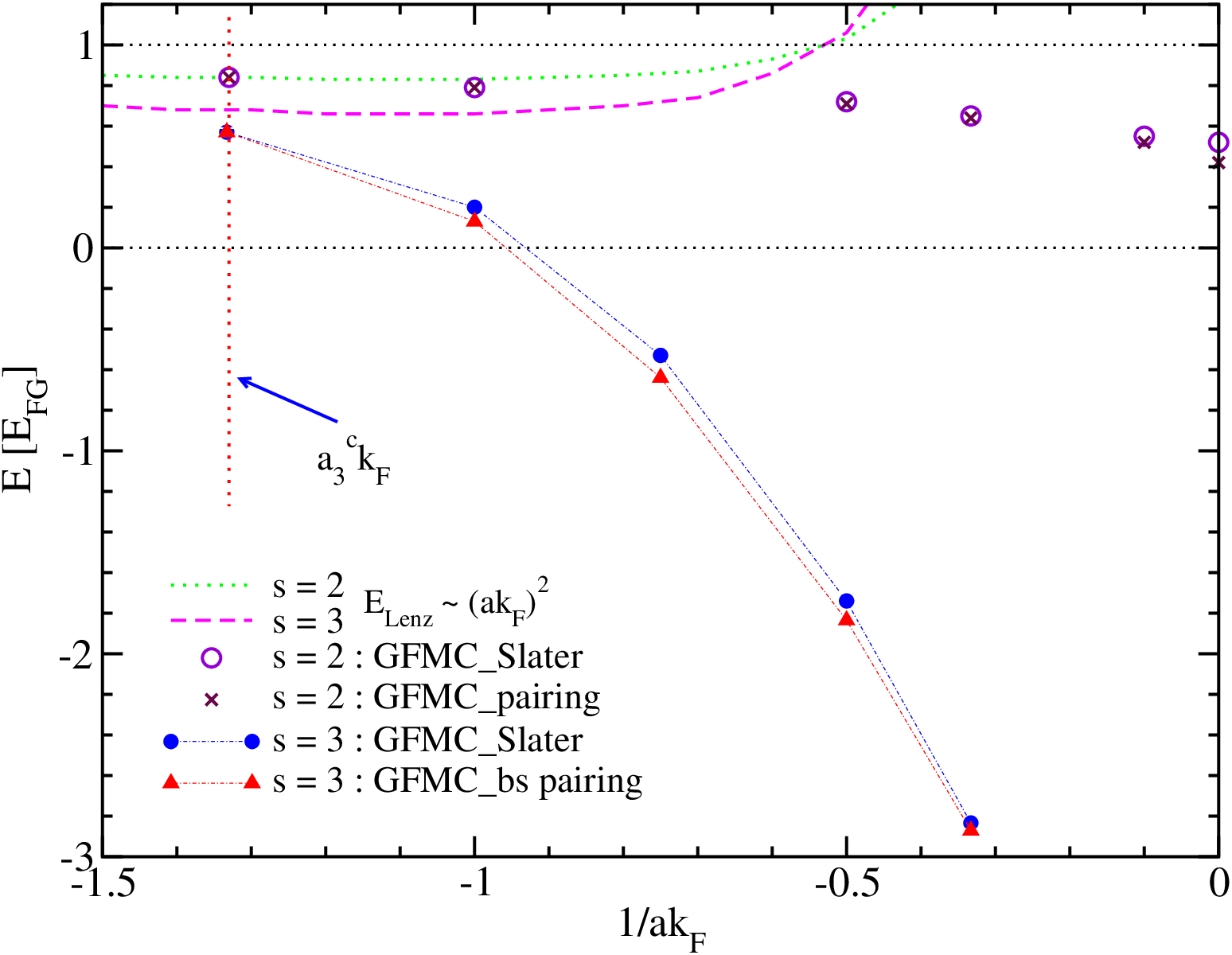}
\caption{(color online) Comparison of the degeneracy two ($s = 2$) and three ($s = 3$) results for $R k_F = 0.32$ in both cases.
 $E_{Lenz}$ (Eq. \ref{eqn_lenz}) gives a good estimate of the normal state energy for $s = 2$ and $1/ak_F \le -1$. 
However for $s = 3$, the match is poor with the GFMC results as the
$R$ dependent terms and three particle effects predominate. 
In the region $1/ak_F \lesssim -0.3$, the SU(2)$\otimes$U(1) broken symmetry pairing state is shown to be the ground state for $s = 3$ Fermi
 gas. However, for $-0.3 \lesssim 1/ak_F$ the SU(3) symmetry is restored(see also Fig. \ref{fig_four}) and the superfluid pairing suppressed.
This behavior for $s=3$ is quantitatively different from the $s=2$ case, where with the increasing interaction
strength (increasing $1/ak_F$), the superfluid pairing state (crosses) is always the favored ground state compared to the
state without superfluid correlation (circles). The size of the symbols approximately correspond to the error bars. }
\label{fig_one}
\end{center}
\end{figure} 
 
\begin{figure}
\begin{center}
\includegraphics[width=8.0cm]{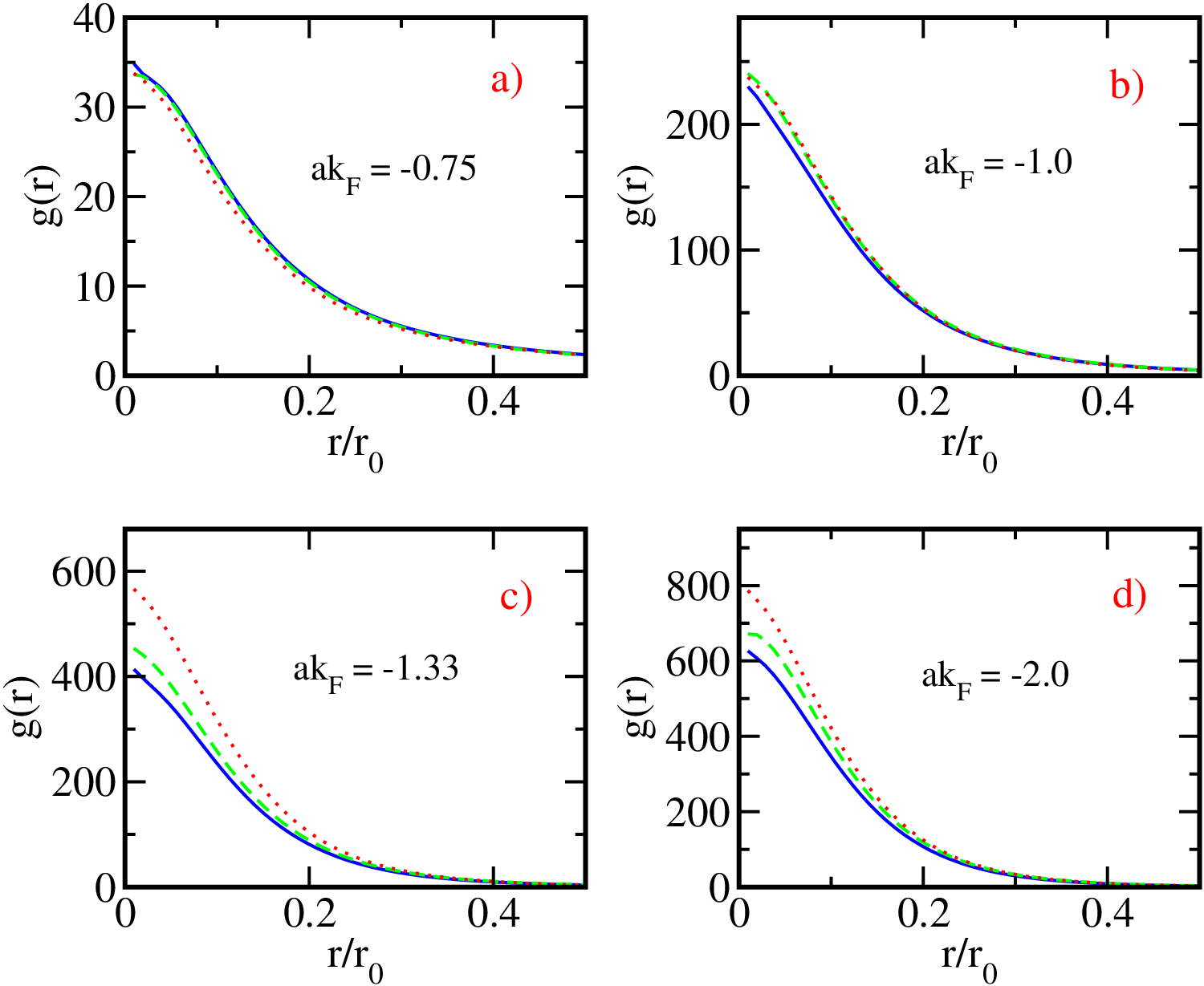}
\caption{(color online) Pair distribution functions at $ak_F = -0.75$(a), $-1.0$(b), $-1.\dot{3}$(c) and $-2.0$(d).
We define pair distributions $g_{\alpha \beta}(r) \sim \langle \sum\limits_{i_\alpha,j_{\beta}} \delta(r - r_{i_\alpha,j_{\beta}} ) \rangle$. 
The angled brackets imply taking thermal average of the possible configurations.  $g_{\alpha \beta}(r)$ has boundary condition $\lim\limits_{r \rightarrow \infty} g_{\alpha \beta}(r) =1$. 
The line corresponds to $g_{GR}(r)$, that is, the pair distribution between the green and red species while the dashed line corresponds
to the distribution between the green and blue species $g_{GB}(r)$. Both $g_{GR}(r)$ and $g_{GB}(r)$ assume
SU(2)$\otimes$U(1) symmetric state (Eq. \ref{eqn_res3_bcs}).
The dotted line represents the pair distribution without
superfluidity $g_{Slater}(r)$ calculated with the SU(3) symmetric Slater wave function (Eq. \ref{eqn_res3_slater}). 
The $g_{Slater}(r)$ is independent of the colors as long as they are different.  All $g(r)$'s are calculated with the optimized $f(r)$ functions.
In the plots c) and d) where the superfluid pairing is strong,  $g_{GR}(0) < g_{GB}(0)  = g_{RB}(0) < g_{Slater}(0)$. 
Thus, on the average the particles that have superfluid pairing correlation remain further apart than the particles that
have no superfluid pairing correlations.  }
\label{fig_two}
\end{center}
\end{figure} 

\begin{figure}
\begin{center}
\includegraphics[width=6.0cm]{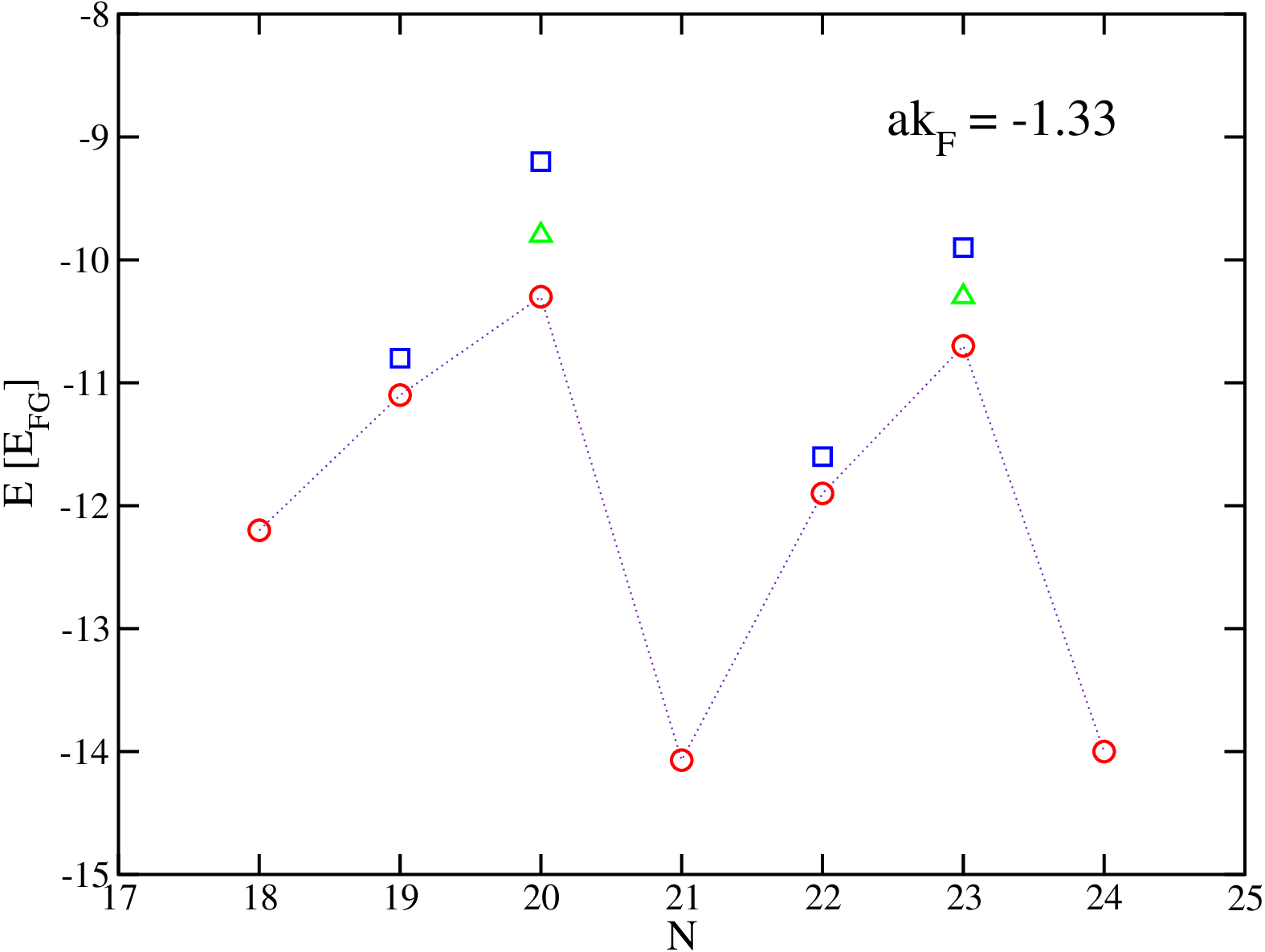}
\caption{(color online) Energy excitations with unbalanced particle numbers (allows broken pairs) at  $ak_F = -1.\dot{3}$. The circles correspond to the set of particle numbers 
 $\{N_G, N_R, N_B \}=$ $\{6,6,6\}$,$\{7,6,6\}$,$\{7,7,6\}$, $\{7,7,7\}$,$\{8,7,7\}$,$\{8,8,7\}$, and $\{8,8,8\}$. 
The squares are $\{6,6,7\}$, $\{6,6,8\}$, $\{7,7,8\}$, and $\{7,7,9\}$ cases. The triangles represent the configurations $\{7,6,7\}$ and $\{8,7,8\}$. 
The size of the symbols approximately correspond to the error bars.}
\label{fig_three}
\end{center}
\end{figure}  
  
\begin{figure}
\begin{center}
\includegraphics[width=8.0cm]{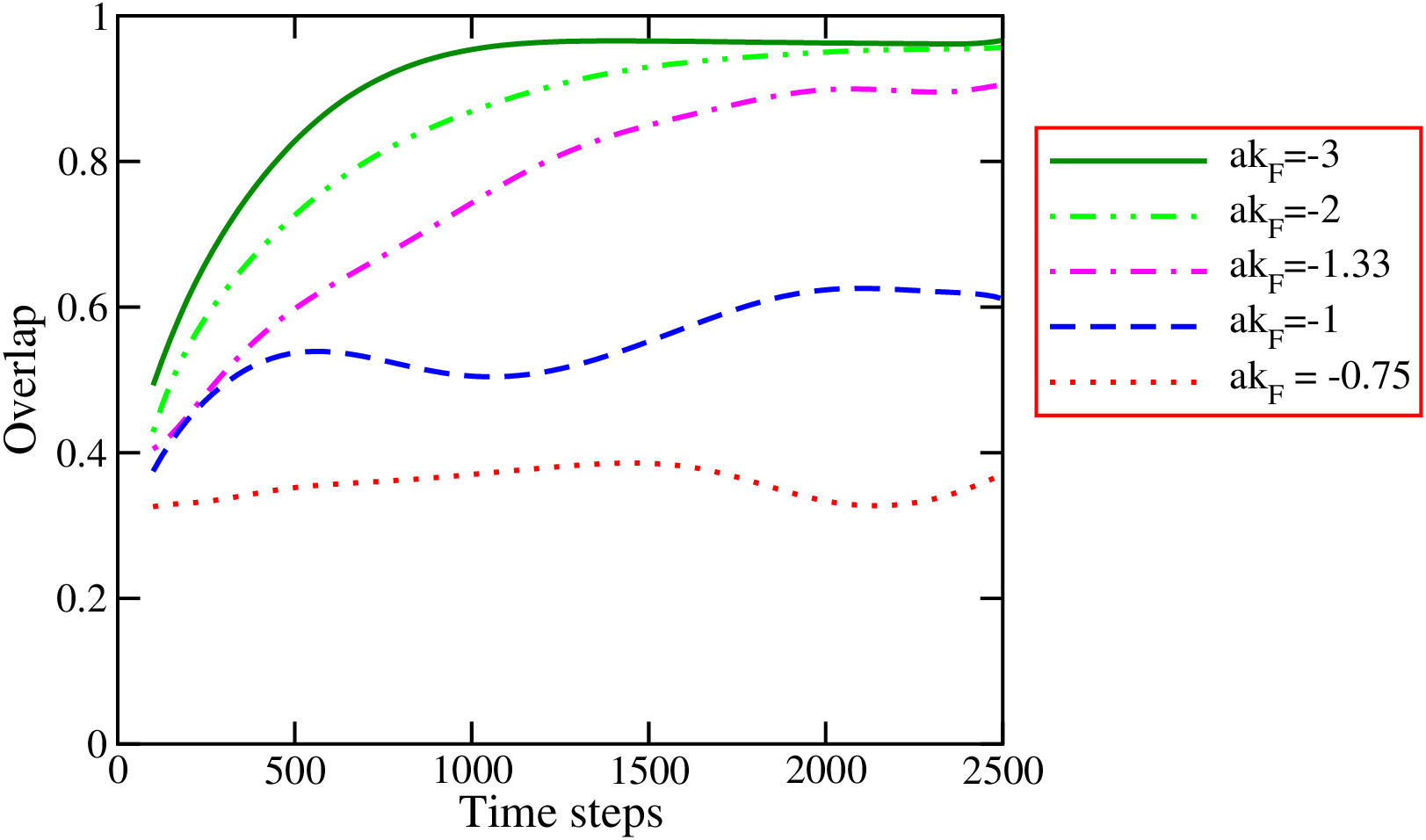}
\caption{(color online) Imaginary time evolution of the overlap of the wave functions defined as $\langle 2|e^{-{\cal H}\tau} |1 \rangle = 
\langle3 | e^{-{\cal H}\tau} |1 \rangle \sim \langle \Psi_{FG} | e^{-{\cal H}\tau} |1 \rangle$ are shown. Each time step is $\delta\tau = 3.6\times10^{-4} \frac{\hbar}{E_{FG}}$. 
The states $|1\rangle$, $|2\rangle$ and $|3\rangle$ are as defined in Eq. \ref{eqn_degens} while $|\Psi_{FG}\rangle$ is as defined in Eq. 
\ref{eqn_res3_slater}. We can see that as the interaction strength is increased, the symmetry is restored with the overlaps
approaching 1.}
\label{fig_four}
\end{center}
\end{figure}

The results of three component Fermi gas energies are summarized in the Fig. \ref{fig_one}. The pairing correlations produce
noticeable effects at $-1 \le 1/ak_F \le -0.5$ (compare the triangles with the circles of the same figure). Close to the $a^c_3$, 
 the pairing effects are small. Here, the energy of the SU(2)$\otimes$U(1) broken symmetry state
 is not distinguishable within the error bars from that of the SU(3) symmetric state (Slater) .
 The system is found stable in the regime of interaction considered in this work
 ($-1.3 \lesssim 1/ak_F \lesssim -0.3$ and $R k_F = 0.32$).

 We also found that it is possible to see the effects of pairing in the $g(r)$'s. 
In the plot a) of Fig. \ref{fig_two} ($a k_F = -0.75$), no difference can be seen in the $g(r)$'s because the pairing is weak. 
However, where the pairing is relevant (in terms of the energy) we notice that $g_{GR}(0) < g_{GB}(0) = g_{RB}(0) < g_{Slater}(0)$
 (where $g_{Slater}(r)$ is the pair distribution obtained with the non-pairing Slater wave function) and the symmetry is broken
 (see the plots c) and d) of the Fig \ref{fig_two}). Green and red particles have superfluid pairing correlations,
 so one may have naively expected $g_{GR}(0) > g_{GB}(0)  = g_{RB}(0)$, but the opposite is found to be true. 
According to this, the relative distances satisfy on average: $r_{GB} = r_{RB} < r_{GR}$. 
The interpretation we can give is that once a Cooper pair is formed, the third particle feels much stronger attraction
 toward the center of the mass of the Cooper pair.  In fact, we can approximately estimate that the potential strength between the center of mass (CM)
 of green-red pair and the blue particle is enhanced from $v_0$ to  $\frac{4}{3} v_0$, while green and red particles interact with the strength $v_0$. 
This comes from the observation that the zero energy scattering equation between the CM of green-red and blue is
$-\frac{\hbar^2}{2m_r}u''(r) + v_0 V_R(r) u(r) = 0$ with $m_r = \frac{2}{3}m$. From the plot d) of the Fig. \ref{fig_two}, it is still not conclusive whether the symmetry could be restored
($g_{GR}(0) = g_{GB}(0) = g_{RB}(0)$) in the regime of strong three particle correlation $-0.5 < 1/ak_F$. In that regime,
the analysis is hindered on the practical ground: it becomes increasingly harder to obtain reliable estimates of $g(r)$
keeping the statistical errors small. However, the Fig. \ref{fig_one} and Fig. \ref{fig_four} show clearly the tendency toward the
restoration of the SU(3) symmetry entering into a regime completely dominated by three particle physics with negligible
paring correlations.

 Using the GFMC technique, energy gaps can be calculated by allowing variations in the numbers of the green and red particles 
while keeping the number of blue particles constant. In this way, we allow breaking of the superfluid pairs.
 We can consider sets of $\{N_G, N_R, N_B \}=$ $\{6,6,6\}$,$\{7,6,6\}$,$\{7,7,6\}$,
 and $\{7,7,7\}$,$\{8,7,7\}$,$\{8,8,7\}$. We observe the usual odd-even staggering of the ground state energy (Fig. \ref{fig_three}).
Furthermore, we can allow the variations $6 \le N_G \le 8$, $6 \le N_R \le 8$, and $6 \le N_B \le 8$ and calculate
the excitation energies. The lowest energy excitations at $ak_F = -1$ and $-1.\dot{3}$ are found with the momentum ${\bf k } = {\bf 0}$
quasiparticle.  In the case of $s = 2$, quasiparticles with zero momentum produce the minimum excitation 
in the $1/ak_F > 0$ (BEC) regime. Thus, this is consistent with the interpretation that the studied regime ($-1.3 \lesssim 1/ak_F \lesssim  -0.3$)
is that of strong three particle correlation.  The energies for the broken pair states are shown in the Fig. \ref{fig_three} for the $ak_F = -1.\dot{3}$ case. 
The gap $\Delta$ is estimated from the data sets represented by circles with total $N = 18,19,20$ and $N = 21,22,23$ respectively.
 The calculated $\Delta/E_{FG} \approx 0.3(3)$.  The error bars are large for the pairing gap since the three particle 
effects predominates rather than two particle pairing. In the Fig. \ref{fig_three}, we notice that the energy dips 
when total $N$ is a multiple of $3$. We interpret this as an effect analogous to that observed in the $g(r)$'s. This is the evidence 
that the trimer interaction that brings the green-red pair and the blue particle together is much stronger than simple pairwise interaction. 
Thus, completing green-red-blue trimer is energetically more favorable than unbalanced excess of one or two species. 
In fact, the trimer binding energy is so strong that $\Delta < |E_{trimer}/3|$
in contrast to $\Delta  \approx |E_{pair}/2|$ of $s=2$ Fermi gas in the BEC regime.
Consequently, in the quasiparticle spectrum, it is expected that we can observe two distinctive {\it gaps}; one due to the
superfluid pairing and another due to the trimer binding. We also notice that at $N = 21$ the dip is as deep as at $N=24$ which 
indicates possible shell closure effect.  As seen in the ground state energy (Fig. \ref{fig_one}) and the overlap of the
wave functions (Fig. \ref{fig_four}), the broken symmetry superfluidity is suppressed in the $ak_F > -0.5$ regime where
the SU(3) symmetry is restored.

For the comparison purpose only, we {\it naively} consider extension of the mean field (BCS-Leggett) method to the three component Fermi gas.
 For this mean field model, there is no $R$ dependence and the variational ground state is always stable. The two color pairing gap of the degeneracy three superfluid
 is given by the same relation as that of the degeneracy two Fermi gas. While in the original BCS formalism, the chemical potential
 $\mu$ is kept constant and  $\Delta_{BCS}/\epsilon_F = \frac{8}{e^2}e^{\pi/2ak_F}$ we consider the BCS-Leggett \cite{leggett1980}
variational formalism ($\Delta_{BCS-Leggett}$). Here, the chemical potential is changed in order to keep the density constant.
The condensation energy of the system is $E_{cond} = - n_0 \frac{\Delta^2}{2}$ where $n_0$ is the state density ($n_0  \equiv \frac{mk_F \Omega}{2 \pi^2 \hbar^2}$).
Dependence on the degeneracy $s$ is included in the condensation energy per particle $\frac{E_{cond}}{N} = -\frac{9}{20 s} \frac{\Delta^2}{E_{FG}}$
because of the relation $\frac{6\pi^2}{s}\rho = k_F^3$. Thus,  at $1/ak_F = -1$, we estimate that $\Delta_{BCS-Leggett}/E_{FG} = 0.33$ and  $E_{cond}/(N E_{FG}) \approx -0.016$.
At  $1/ak_F = -0.75$ (or $ak_F = 1.\dot{3}$) ,we have $\Delta_{BCS-Leggett}/E_{FG} = 0.5$ and  $E_{cond}/(N E_{FG}) \approx -0.038$.
This estimate is close to the one calculated by GFMC (Fig. \ref{fig_three},  $\Delta/E_{FG} \sim 0.3$ ) at the same interaction strength. 
 
 Although for $s=3$ Fermi gas, $R$ dependence cannot be removed, clear qualitative differences between the $s = 2$ and $s=3$ Fermi gases emerge.
 Unlike in the $s=2$ Fermi gas where both the superfluid pair and the bound state are qualitatively similar, in the $s=3$ Fermi gas the paired states
decouple from the bound state(trimer) in energy. This can be clearly observed in the quasiparticle excitation spectrum. 
Realistic interaction potential and channel dependence of the interactions are necessary in order to
 produce not only qualitative but also quantitatively correct results for a given three component Fermi gas. We found a regime of interaction strength
where the broken symmetry pairing is clearly detectable, beyond which three particle effects dominate and the symmetry is restored. 
Also, non trivial dependence of the $g(r)$'s on the pairing correlations was discussed. 
This work has been supported in part by the US National Science Foundation via grant PHY 00-98353 and PHY 03-55014. SYC also 
acknowledges support by the DARPA grant BAA 06-19. The authors acknowledge useful comments from J. Carlson and A. Bulgac. One of the authors(VRP) passed away during the preparation
 of the present manuscript and the work was posthumously completed.


\begin{thebibliography}{16}
\expandafter\ifx\csname natexlab\endcsname\relax\def\natexlab#1{#1}\fi
\expandafter\ifx\csname bibnamefont\endcsname\relax
  \def\bibnamefont#1{#1}\fi
\expandafter\ifx\csname bibfnamefont\endcsname\relax
  \def\bibfnamefont#1{#1}\fi
\expandafter\ifx\csname citenamefont\endcsname\relax
  \def\citenamefont#1{#1}\fi
\expandafter\ifx\csname url\endcsname\relax
  \def\url#1{\texttt{#1}}\fi
\expandafter\ifx\csname urlprefix\endcsname\relax\def\urlprefix{URL }\fi
\providecommand{\bibinfo}[2]{#2}
\providecommand{\eprint}[2][]{\url{#2}}


\bibitem{giorgini2007} See for example summaries by S.~Giorgini, L.~P.~Pitaevskii and S.~Stringari, Rev. Mod.
Phys. {\bf 80}, 1215 (2007), and by R.~Grimm  cond-mat/0703091 (2007). 

\bibitem{ketterle2008} W.~Ketterle, and M.~W.~Zwierlein, arXiv:0801.2500v1 (2008). 

\bibitem{lenz1929} W.~Lenz, Z. Physik {\bf 56}, 778 (1929).   

\bibitem{huang1957} K.~Huang, and C.~N.~Yang, Phys. Rev. {\bf 105}, 767 (1957).     

\bibitem{galitskii1958} V.~M.~Galitskii, Sov. Phys. JETP {\bf 7}, 104 (1958).  
 

\bibitem{baker1999} G.~A. Baker, Phys. Rev. C {\bf 60}, 054311 (1999).  

\bibitem{bulgac2002} A.~Bulgac, Phys. Rev. Lett. {\bf 89}, 050402 (2002).  

\bibitem{carlson2003} J.~Carlson, S.~Y.~Chang, V.~R.~Pandharipande, and K.~E.~Schmidt, 
Phys. Rev. Lett. {\bf 91}, 50401 (2003).  

\bibitem{carlson2008} J. Carlson, and S.~Reddy, Phys. Rev. Lett. Phys. Rev. Lett. {\bf 100}, 150403 (2008).

\bibitem{astra2004} G.~E.~Astrakharchik, J.~Boronat, J.~Casulleras, and S.~Giorgini, Phys. Rev. Lett. {\bf 93}, 200404 (2004).  
  
\bibitem{chang2004} S.~Y.~Chang, V.~R.~Pandharipande, J.~Carlson, and K.~E.~Schmidt, Phys. Rev. A. {\bf 70}, 043602 (2004).  

\bibitem{honerkamp2004} C.~Honerkamp, and W.~Hofstetter, Phys. Rev. B {\bf 70}, 094521 (2004).  

\bibitem{torma2006} T.~Paananen, J.~P. Martikainen, and P.~T\"orm\"a, Phys. Rev. A {\bf 73}, 053606 (2006).  

\bibitem{torma2007} T.~Paananen, P.~T\"orm\"a, and J.~P. Martikainen, Phys. Rev. A {\bf 75}, 023622 (2007).


%\bibitem{bedaque2006} P.~F.~Bedaque, and J.~P.~D'Incao,  cond-mat/0602525 (2006).

\bibitem{jochim2008} T.~B.~Ottensten,~T.~Lompe, ~M.~Kohnen,~A.~N.~Wenz,~and S.~Jochim, Phys. Rev. Lett.
{\bf 101}, 203202 (2008).

\bibitem{gupta2003} S.~Gupta, Z.~Hadzibabic, M.~W.~Zwierlein, C.~A.~Stan, K.~Dieckmann,  
C.~H.~Schunck, E.~G.~M.~van Kempen, B.~J.~Verhaar, and W.~Ketterle,
Science {\bf 300}, 1723 (2003).

\bibitem{barter2005} M.~Bartenstein, A.~Altmeyer, S.~Riedl, R.~Geursen, S.~Jochim, C.~Chin, J.~H.~Denschlag,
 R.~Grimm, A.~Simoni, E.~Tiesinga, C.~J.~Williams, and P.~S.~Julienne, Phys. Rev. Lett. {\bf 94}, 103201 (2005).

\bibitem{alford2001} M.~Alford, Ann. Rev. Nucl. Part. Sci. {\bf 51}, 131 (2001).


\bibitem{modawi1997} A.~G.~K.~Modawi, and A.~J.~Leggett, Journal of Low Temp. Phys. {\bf 109}, 625 (1997).

\bibitem{thomas1935} L.~H.~Thomas, Phys. Review {\bf 47}, 903 (1935).

\bibitem{efimov70} V.~Efimov, Phys. Lett. {\bf 33B}, 563 (1970).
 
\bibitem{efimov71} V.~Efimov, Sov. J. Nucl. Phys. {\bf 12}, 589 (1971). 
 
\bibitem{lim77} T.~K.~Lim, K.~Duffy, and W.~Damert, Phys. Rev. Lett. {\bf 38}, 341 (1977).

\bibitem{esry2005} J.~P.~D'Incao, and  B.~D.~Esry, Phys. Rev. A {\bf 72}, 032710 (2005).
 
\bibitem{chang2005} S.~Y.~Chang, and V.~R.~Pandharipande, Phys. Rev. Lett. {\bf 95}, 080402 (2005).  
  
\bibitem{leggett1980} A.~J.~ Leggett, {\it Modern Trends in the Theory of Condensed Matter},
  edited by A.~Pekalski, and R.~ Przystawa (Springer-Verlag, Berlin, 1980).  
  
\end{thebibliography}
\end{document}